\documentclass[runningheads]{llncs}
\usepackage{thm-restate}
\usepackage{amsfonts}
\usepackage[page,header]{appendix}
\usepackage{titletoc}
\usepackage{setspace}
\usepackage{enumitem}
\usepackage{wrapfig}
\usepackage[framemethod=TikZ]{mdframed}
\usepackage[T1]{fontenc}
\usepackage[utf8]{inputenc} %support umlauts in the input
\usepackage{subcaption}
\usepackage{etoolbox}
\usepackage{xifthen}
\usepackage{xspace}
\usepackage{thmtools} 
\usepackage{thm-restate}

\usepackage{stmaryrd}
\SetSymbolFont{stmry}{bold}{U}{stmry}{m}{n}
\usepackage{anyfontsize}

\usepackage{proof}
\usepackage{mathtools} %% for \mathrlap{}
\usepackage{mathrsfs}
\usepackage[most]{tcolorbox}
\usepackage{graphicx,graphics}
\usepackage{url}
\usepackage{xcolor}
\usepackage{comment}
\usepackage{rotating}
\usepackage{listings}
\usepackage{multicol}
\usepackage{multirow}
\usepackage[noend]{algpseudocode}
\usepackage{tikz}
\usepackage{fancybox}
\usetikzlibrary{positioning,calc,backgrounds,fit,shapes,arrows.meta,cd}
\usetikzlibrary{arrows,shapes,automata,spy,positioning,
  calc,fit,arrows.meta,patterns,backgrounds,decorations.pathreplacing,shadows}  

\usepackage{import}
\usepackage{makecell}
\usepackage{relsize}
\usepackage{listings}
\usepackage{csquotes}
\usepackage{booktabs}
\usepackage[group-four-digits,per-mode=fraction]{siunitx}
\usepackage{adjustbox} %% for resizing tables
\usepackage{arydshln} %% dashed \hline
\usepackage{graphicx}

\usepackage[final]{fixme} % add 'inline' and/or 'nomargin' if required
\fxsetup{theme=color,mode=multiuser}
\FXRegisterAuthor{JL}{aJL}{JL}
\FXRegisterAuthor{KI}{aKI}{KI}
\FXRegisterAuthor{RN}{aRN}{RN}

\usepackage{upquote}
\makeatletter
\RequirePackage[bookmarks,unicode,colorlinks=true]{hyperref}\def\@citecolor{blue}\def\@urlcolor{blue}\def\@linkcolor{blue}
\usepackage[capitalise,nameinlink,noabbrev]{cleveref}
\def\orcidID#1{\smash{\href{http://orcid.org/#1}{\protect\raisebox{-1.25pt}{\protect\includegraphics{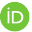}}}}}
\makeatother
\newcommand{\PSEND}[2]{\ptp{#1} ! \msg{#2}}
\newcommand{\PRECEIVE}[2]{\ptp{#1} ? \msg{#2}}
\newcommand{\ptp}[1]{\texttt{#1}}
\newcommand{\msg}[1]{\textit{#1}}

\tikzset{
  every state/.style={minimum size=1pt,inner sep=1.2pt, initial text={}},
  mycfsm/.style={
    font=\scriptsize,
    initial where=left,
    initial distance=0.25cm,
    ->,>=stealth,auto, node distance=0.8cm and 0.8cm,
    scale=1, every node/.style={transform shape},
    baseline=(current  bounding  box.center)
  },
  line/.style = {draw,->, rounded corners=0.07cm,>=latex},
  deliv/.style = {draw, rectangle, rounded corners, fill=white,drop shadow,align=center},
  manip/.style = {draw, ellipse, fill=white,drop shadow,align=center,inner sep=1pt},
  every picture/.style={/utils/exec={\sffamily}}
}

\newcommand{\kmc}{$k$-MC\xspace} 
\newcommand{\kmclib}{{\tt kmclib}\xspace}

\definecolor{ForestGreen}{RGB}{34,139,34}
\newcommand{\chan}{\textit{ch}}

\newcommand{\myparagraph}[1]{\smallskip \noindent \textbf{#1.}\ }

\begin{document}
%
% \title{All-in-one: Safe Concurrent Programming in OCaml}
%\title{\kmclib{}: Bounded Verification of Session Types Meets Meta-Programming in OCaml}
\title{\texttt{Kmclib}: Automated Inference and Verification of Session Types from OCaml Programs}
\titlerunning{\texttt{Kmclib}: Automated Inference and Verification of Session Types}

\author{
  Keigo Imai\inst{1}\orcidID{0000-0003-1602-8473}
  \and
  Julien Lange\inst{2}\orcidID{0000-0001-9697-1378}
  \and
  Rumyana Neykova\inst{3}\orcidID{0000-0002-2755-7728}
}

\institute{Gifu University, Japan
  \and Royal Holloway, University of London, UK
  \and Brunel University London, UK  
}

%
%\titlerunning{Abbreviated paper title}
% If the paper title is too long for the running head, you can set
% an abbreviated paper title here
%
% \author{First Author\inst{1}\orcidID{0000-1111-2222-3333} \and
% Second Author\inst{2,3}\orcidID{1111-2222-3333-4444} \and
% Third Author\inst{3}\orcidID{2222--3333-4444-5555}}
%
% \authorrunning{F. Author et al.}
% First names are abbreviated in the running head.
% If there are more than two authors, 'et al.' is used.
%
%\institute{Princeton University, Princeton NJ 08544, USA \and
%Springer Heidelberg, Tiergartenstr. 17, 69121 Heidelberg, Germany
%\email{lncs@springer.com}\\
%\url{http://www.springer.com/gp/computer-science/lncs} \and
%ABC Institute, Rupert-Karls-University Heidelberg, Heidelberg, Germany\\
%\email{\{abc,lncs\}@uni-heidelberg.de}}
%
\maketitle              % typeset the header of the contribution
\begin{abstract}
  Theories and tools based on multiparty session types offer
  correctness guarantees for concurrent programs that communicate
  using message-passing.
  These guarantees usually come at the cost of an intrinsically
  top-down approach,
  %  that does not fit well with modern development
  % practices and 
  which requires the communication behaviour of the entire program to
  be specified as a global type.
  
  This paper introduces \kmclib: an OCaml library that supports the
  development of \emph{correct} message-passing programs without 
  having to write any types.
  The library utilises the meta-programming facilities of OCaml to
  automatically infer the session types of concurrent programs and
  verify their compatibility (\kmc~\cite{LangeY19}).
  Well-typed programs, written with \kmclib, do not lead to
  communication errors and cannot get stuck.
  % We leverage the {\em type-aware} macro system of OCaml
  % and recent abvances in session types compatibility checking. 
  
  %
  % These programs are then statically verified using the power of the
  % {\em type-aware} macro system of OCaml and recent advances in
  % session types to deliver strong correctness guarantees.
  % 
\end{abstract}

\newcommand{\must}{\ourlibrary}
\newcommand{\mustL}{{\sf{MiO}}\xspace}
\newcommand{\globalcombinators}{global combinators}

\definecolor{dkblue}{rgb}{0,0.1,0.5}
\definecolor{dkgreen}{rgb}{0,0.4,0.1}
\definecolor{dkred}{rgb}{0.4,0,0}

\newcommand{\highlightcolor}{orange!30}
\definecolor{linkColor}{rgb}{0,0,0.5}
\definecolor{pblue}{rgb}{0.13,0.13,1}
\definecolor{purple}{rgb}{0.5,0,0.5}
\definecolor{beige}{rgb}{0,0.5,0.5}
\definecolor{pred}{rgb}{0.9,0,0}
\definecolor{WhiteSmoke}{rgb}{0.96, 0.96, 0.96}
\definecolor{mygray}{gray}{0.6}

\newcommand{\CODESIZETINY}{\footnotesize}
\newcommand{\CODESIZE}{\small}  % In llncs footnotesize = small
\newcommand{\CODESIZENORMAL}{\small}  % In llncs footnotesize = small
\newcommand{\LSTCODESIZE}{\fontsize{8pt}{9.5pt}}  % skip ~ 1.2*size
\newcommand{\CODESTYLE}{\ttfamily}
\newcommand{\CODE}[1]{{\CODESIZE\CODESTYLE #1}}
\newcommand{\MCODE}[1]{\textnormal{\small\CODESTYLE{#1}}}
\newcommand{\CODEKW}[1]{{\color{purple}\CODESIZE\CODESTYLE #1}}

\newcommand{\CODECONST}[1]{{\color{dkblue}\CODESTYLE\CODESIZE\itshape#1}}

\newcommand{\TICK}{\color{dkgreen}{\ding{51}}}
\newcommand{\CROSS}{\color{red}{\ding{55}}}

\newcommand{\WARNING}{\color{orange}{\ding{115}\hspace{-6pt}\color{black}{\small\textbf{!}}}}
\newcommand{\QUESTION}{{\normalsize\color{BurntOrange}\textbf{?}}}

\mdfdefinestyle{cdescription}{%
  hidealllines=true,%no lines drawn
  %backgroundcolor= gray!30,
  %innermargin =-1cm,
   %outermargin =-1cm,
  skipbelow= \baselineskip,
  roundcorner= 2pt,
  splittopskip=\topskip,skipbelow=\baselineskip,%
   skipabove=\baselineskip
}

\mdfdefinestyle{adescription}{%
  hidealllines=true,%no lines drawn
  backgroundcolor= red!15,
   innermargin =-1cm,
   outermargin =-1cm,
  skipbelow= \baselineskip,
  roundcorner= 2pt,
  splittopskip=\topskip,skipbelow=\baselineskip,%
   skipabove=\baselineskip
}

\surroundwithmdframed[style=adescription]{adescription}

\newenvironment{myenv}[1][]
  {\begin{mdframed}[style=cdescription,font=\scriptsize,#1]\begin{tabbing}}
  {\end{tabbing}\end{mdframed}}

\begin{comment}
\newcommand{\ERROR}[4]%
{%
\begin{textblock*}{#1}(#2,#3)
\begin{tikzpicture}
\node[inner sep=2pt,rectangle,draw,color=gray,text=black,font=\ssmall,fill=yellow!35] (BOX) {\hspace{8pt} #4~~};
\node[anchor=west,xshift=-2pt] at (BOX.west) {\includegraphics[scale=0.8]{error.png}};
\end{tikzpicture}
\end{textblock*}
}
\end{comment}

\lstdefinelanguage{SOCaml}{
    morekeywords= {val, let, new, lazy, match, with, rec, open, module, namespace, type, of, member, and, for, while, true, false, in, do, fun, return, yield, try, mutable, if, then, else, cloud, async, static, use, abstract, interface, inherit, finally, Thread, begin, end},
	otherkeywords={ let!, return!, do!, yield!, use!, var, from, select, where, order, by, match, goto},
    sensitive=false,
	morekeywords = [3]{choice_at, finish, fix, out, inp, dlin, channel, send, receive, close},
    keywordstyle = [3]\color{purple},
    morekeywords = [4]{c, a, s},
    morecomment=[s][\color{dkgreen}]{{(*}{*)}},
    moredelim=[is][\color{gray}]{--!}{!--},
    moredelim=[is][\color{dkred}]{--?}{?--},
    moredelim=**[is][\btHL]{£}{£},
    moredelim=**[is][\bfseries\color{dkred}]{&}{&},
    %% moredelim=**[is][\btHL]{~}{~},
    moredelim=**[is][\berrHL]{€}{€},
    morestring=[b]",
    stringstyle=\color{dkred}
}

\lstset{
	columns=flexible, %% mandatory
	captionpos=b,
	escapeinside={(*}{*)},
	float=hbp,
	frame=none,
	mathescape=true,
	numbers=none,
	numberstyle=\tiny,
	showspaces=false,
	showstringspaces=false,
	showtabs=false,
	tabsize=1,
	commentstyle     = \color{Green},      % comment color
    keywordstyle=\CODESIZENORMAL\CODESTYLE\color{dkblue},
    stringstyle      = \color{DarkRed},    % string color
    framesep         = 3pt,   % expand outward
    %% OCaml by default
    language = SOCaml,
    alsoletter = {_},
    basicstyle=\CODESIZENORMAL\CODESTYLE,
    emph={string,str,int, real, bool, unit},
    emphstyle=\color{dkred},
    breaklines=true,
    backgroundcolor = {},
    literate=
    {-->}{{\color{purple}-{}->}}1
    %% {-->}{$\color{purple}{\longrightarrow}$}1
    {:?}{$\color{dkblue}{:? }$}1
    {<}{<}1
     {(}{(}1
    {\%}{{\color{blue}\%}}1
     {@@}{{\color{dkblue}@@}}1
    {<@}{{\color{blue}<@} \quad}1
    {@>}{\quad{\color{blue}@}>}1
    {>}{>}1
    {._}{{$\,$}}1
    {,,}{{$\,\elipc\,$}}1
    {**}{{,}}1
}

%% reset OCaml-specific settings
\lstdefinestyle{BASE}{
  mathescape=false,
  keywordstyle=\CODESIZENORMAL\CODESTYLE\color{dkblue}, % keyword color
  alsoletter={},
  basicstyle={},
  emph={},
  emphstyle={},
  breaklines=false,
  literate={},
}

%\lstMakeShortInline[columns=fixed]|
\newcommand{\SCRIB}[1]{\lstinline[style=SCRIB]+#1+}
\newcommand{\SCRIBAPI}[1]{\lstinline[style=SCRIBAPI]+#1+}

\lstdefinestyle{SCRIBBASE}%
{%
	style=BASE,
	language=Java,
	morekeywords=[1]{
		protocol, role, choice, at, or, from, to, rec, par, and, interrupt, by, finish, continue, global, local, self, interruptible, with, as, catches, type, sig, aux, explicit, connect, wrap, disconnect, accept, wrapClient, wrapServer},
	%morekeywords=[2]{int,String,Date},
	%literate={>=}{$\geq\ $}{2}{<=}{$\leq\ $}{2}
	%moredelim=*[s][\LSTFONTSIZE\color{dkgreen}]{<}{>}
  %moredelim=**[is][\only<+>{\color{black}\lstset{style=SCRIB}}]{@+}{@}
	%% FIXME: the \lstset doesn't actually work
  %moredelim=**[is][\only<1->{\color{red}\lstset{style=ERROR}}]{@!}{@},
  moredelim=**[is][{\color{red}\lstset{style=ERROR}}]{@!}{@},
  moredelim=**[is][\only<1>{\color{red}\lstset{style=ERROR}}]{@!1}{@},
  moredelim=**[is][\only<2>{\color{red}\lstset{style=ERROR}}]{@!2}{@},
  moredelim=**[is][\only<3>{\color{red}\lstset{style=ERROR}}]{@!3}{@},
  moredelim=**[is][\only<4>{\color{red}\lstset{style=ERROR}}]{@!4}{@},
  moredelim=**[is][\only<5>{\color{red}\lstset{style=ERROR}}]{@!5}{@},
  %moredelim=**[is][\only<1->{\color{dkgreen}\lstset{style=ASSIST}}]{@?}{@},
  moredelim=**[is][{\color{dkgreen}\lstset{style=ASSIST}}]{@?}{@},
  moredelim=**[is][\only<1>{\color{dkgreen}\lstset{style=ASSIST}}]{@?1}{@},
  moredelim=**[is][\only<2>{\color{dkgreen}\lstset{style=ASSIST}}]{@?2}{@},
  moredelim=**[is][\only<3>{\color{dkgreen}\lstset{style=ASSIST}}]{@?3}{@},
  moredelim=**[is][\only<4>{\color{dkgreen}\lstset{style=ASSIST}}]{@?4}{@},
  moredelim=**[is][\only<5>{\color{dkgreen}\lstset{style=ASSIST}}]{@?5}{@},
  moredelim=**[is][\only<-2>{\color{red}\lstset{style=ERROR}}]{@!-2}{@},
  moredelim=**[is][\only<2->{\color{red}\lstset{style=ERROR}}]{@!2-}{@}
}

\lstdefinestyle{SCRIBBG}%
{%
	style=SCRIBBASE,
	basicstyle=\CODESTYLE\CODESIZE\color{black!40},
  %keywordstyle=\color{red!40},
	keywordstyle=[1]{\color{dkblue!40}},
	keywordstyle=[2]{\color{dkgreen!40}},
  commentstyle=\itshape\color{purple!40},
  %commentstyle=\itshape\color{dkgreen!40},
	identifierstyle=\color{black!40},
	stringstyle=\color{teal!40},
 	emphstyle=\color{dkblue!40}
}

\lstdefinestyle{LSTSCRIBBG}%
{%
	style=SCRIBBG,
	basicstyle=\LSTCODESIZE\CODESTYLE\color{black!40},
  moredelim=**[is][\color{black}\lstset{style=SCRIB}]{@-}{@},
  moredelim=**[is][\only<1>{\color{black}\lstset{style=LSTSCRIB}}]{@1}{@},
  moredelim=**[is][\only<2>{\color{black}\lstset{style=LSTSCRIB}}]{@2}{@},
  moredelim=**[is][\only<3>{\color{black}\lstset{style=LSTSCRIB}}]{@3}{@},
  moredelim=**[is][\only<4>{\color{black}\lstset{style=LSTSCRIB}}]{@4}{@},
  moredelim=**[is][\only<5>{\color{black}\lstset{style=LSTSCRIB}}]{@5}{@},
  moredelim=**[is][\only<6>{\color{black}\lstset{style=LSTSCRIB}}]{@6}{@},
  moredelim=**[is][\only<7>{\color{black}\lstset{style=LSTSCRIB}}]{@7}{@},
  moredelim=**[is][\only<8>{\color{black}\lstset{style=LSTSCRIB}}]{@8}{@},
  moredelim=**[is][\only<9>{\color{black}\lstset{style=LSTSCRIB}}]{@9}{@},
  moredelim=**[is][\only<10>{\color{black}\lstset{style=LSTSCRIB}}]{@10}{@},
  moredelim=**[is][\only<1->{\color{black}\lstset{style=LSTSCRIB}}]{@1-}{@},
  moredelim=**[is][\only<2->{\color{black}\lstset{style=LSTSCRIB}}]{@2-}{@},
  moredelim=**[is][\only<3->{\color{black}\lstset{style=LSTSCRIB}}]{@3-}{@},
  moredelim=**[is][\only<4->{\color{black}\lstset{style=LSTSCRIB}}]{@4-}{@},
  moredelim=**[is][\only<-2>{\color{black}\lstset{style=LSTSCRIB}}]{@-2}{@}
}

\lstdefinestyle{SCRIB}%
{%
	style=SCRIBBASE,
	basicstyle=\CODESTYLE\CODESIZE\color{black},
  %keywordstyle=\color{red},
	keywordstyle=[1]{\color{dkblue}},
	keywordstyle=[2]{\color{dkgreen}},
  commentstyle=\itshape\color{purple},
  %commentstyle=\itshape\color{dkgreen},
	identifierstyle=\color{black},
	stringstyle=\color{teal},
 	emphstyle=\color{dkblue}
}

\lstdefinestyle{LSTSCRIB}%
{%
	style=SCRIB,
	basicstyle=\LSTCODESIZE\CODESTYLE\color{black},
  moredelim=**[is][\only<1>{\color{black!40}\lstset{style=LSTSCRIBBG}}]{@1}{@},
  moredelim=**[is][\only<2>{\color{black!40}\lstset{style=LSTSCRIBBG}}]{@2}{@},
  moredelim=**[is][\only<3>{\color{black!40}\lstset{style=LSTSCRIBBG}}]{@3}{@},
  moredelim=**[is][\only<4>{\color{black!40}\lstset{style=LSTSCRIBBG}}]{@4}{@},
  moredelim=**[is][\only<5>{\color{black!40}\lstset{style=LSTSCRIBBG}}]{@5}{@},
  moredelim=**[is][\only<6>{\color{black!40}\lstset{style=LSTSCRIBBG}}]{@6}{@},
  moredelim=**[is][\only<7>{\color{black!40}\lstset{style=LSTSCRIBBG}}]{@7}{@},
  moredelim=**[is][\only<8>{\color{black!40}\lstset{style=LSTSCRIBBG}}]{@8}{@},
  moredelim=**[is][\only<9>{\color{black!40}\lstset{style=LSTSCRIBBG}}]{@9}{@},
  moredelim=**[is][\only<-2>{\color{black!40}\lstset{style=LSTSCRIBBG}}]{@-2}{@},
  moredelim=**[is][\only<1->{\color{black!40}\lstset{style=LSTSCRIBBG}}]{@1-}{@},
  moredelim=**[is][\only<2->{\color{black!40}\lstset{style=LSTSCRIBBG}}]{@2-}{@},
  moredelim=**[is][\only<3->{\color{black!40}\lstset{style=LSTSCRIBBG}}]{@3-}{@},
  moredelim=**[is][\only<4->{\color{black!40}\lstset{style=LSTSCRIBBG}}]{@4-}{@},
  moredelim=**[is][\only<1-2>{\color{black!40}\lstset{style=LSTSCRIBBG}}]{@1-2}{@}
}

\lstdefinestyle{ERROR}%
{%
	style=SCRIBBASE,
	basicstyle=\CODESTYLE\CODESIZE\color{red},
 % keywordstyle=\color{red},
	keywordstyle=[1]{\color{red}},
	keywordstyle=[2]{\color{red}},
	identifierstyle=\color{red}
}

\lstdefinestyle{ASSIST}%
{%
	style=SCRIBBASE,
	basicstyle=\CODESTYLE\CODESIZE\color{green},
  %keywordstyle=\color{dkgreen},
	keywordstyle=[1]{\color{dkgreen}},
	keywordstyle=[2]{\color{dkgreen}},
	identifierstyle=\color{dkgreen}
}

\lstdefinestyle{CONST}%
{%
	style=SCRIBAPIBASE,
	basicstyle=\CODESTYLE\CODESIZE\color{dkblue},
	keywordstyle=[1]{\color{dkblue}},
	keywordstyle=[2]{\color{dkblue}},
	identifierstyle=\color{dkblue}
}

\lstdefinestyle{SCRIBAPIBASE}%
{%
	language=Java,
	morekeywords=[1]{},
	morekeywords=[2]{connect, accept, send, receive, enterScope, start, end, invite, async, sync, branch, getOp},
	morekeywords=[3]{val},
	%morekeywords=[4]{C, S, Add, Res, Bye},
	basicstyle=\CODESTYLE\CODESIZE\color{black},
	keywordstyle=[1]{\bfseries\color{purple}},
	keywordstyle=[2]{\color{purple}},
	keywordstyle=[3]{\color{dkblue}},
	keywordstyle=[4]{\bfseries\itshape\color{dkblue}},
  commentstyle=\itshape\color{dkgreen},
	%% FIXME: the \lstset doesn't actually work
  moredelim=**[is][\bfseries\itshape\color{dkblue}]{@=}{@},  %% FIXME: not bold
	moredelim=**[is][\only<1>{\bfseries\itshape\color{dkblue}}]{@=1}{@},
  moredelim=**[is][\only<2>{\bfseries\itshape\color{dkblue}}]{@=2}{@},
  moredelim=**[is][\only<3>{\bfseries\itshape\color{dkblue}}]{@=3}{@},
  moredelim=**[is][\only<4>{\bfseries\itshape\color{dkblue}}]{@=4}{@},
  moredelim=**[is][\only<5>{\bfseries\itshape\color{dkblue}}]{@=5}{@},
	moredelim=**[is][\only<1-2>{\bfseries\itshape\color{dkblue}}]{@=1-2}{@},
	moredelim=**[is][\only<4->{\bfseries\itshape\color{dkblue}}]{@=4-}{@},
  moredelim=**[is][\only<1>{\color{red}\lstset{style=ERROR}}]{@!1}{@},
  moredelim=**[is][\only<2>{\color{red}\lstset{style=ERROR}}]{@!2}{@},
  moredelim=**[is][\only<3>{\color{red}\lstset{style=ERROR}}]{@!3}{@},
  moredelim=**[is][\only<4>{\color{red}\lstset{style=ERROR}}]{@!4}{@},
  moredelim=**[is][\only<5>{\color{red}\lstset{style=ERROR}}]{@!5}{@},
  moredelim=**[is][{\color{dkgreen}\lstset{style=ASSIST}}]{@?}{@},
  moredelim=**[is][\only<1>{\color{dkgreen}\lstset{style=ASSIST}}]{@?1}{@},
  moredelim=**[is][\only<2>{\color{dkgreen}\lstset{style=ASSIST}}]{@?2}{@},
  moredelim=**[is][\only<3>{\color{dkgreen}\lstset{style=ASSIST}}]{@?3}{@},
  moredelim=**[is][\only<4>{\color{dkgreen}\lstset{style=ASSIST}}]{@?4}{@},
  moredelim=**[is][\only<5>{\color{dkgreen}\lstset{style=ASSIST}}]{@?5}{@}
}

\lstdefinestyle{SCRIBAPIBG}%
{%
	style=SCRIBAPIBASE,
	basicstyle=\CODESTYLE\CODESIZE\color{black!40},
  keywordstyle=\color{red!40},
	keywordstyle=[1]{\bfseries\color{purple!40}},
	keywordstyle=[2]{\color{purple!40}},
	keywordstyle=[3]{\color{dkblue!40}},
	keywordstyle=[4]{\bfseries\itshape\color{dkblue!40}},
  commentstyle=\color{dkgreen!40},
	%commentstyle=\itshape\color{purple},
	identifierstyle=\color{black!40},
	stringstyle=\color{teal!40},
 	emphstyle=\color{dkblue!40}
}

\lstdefinestyle{LSTSCRIBAPIBG}%
{%
	style=SCRIBAPIBG,
	basicstyle=\LSTCODESIZE\CODESTYLE\color{black!40},
  moredelim=**[is][\color{black}\lstset{style=LSTSCRIBAPI}]{@-}{@},
  moredelim=**[is][\only<1>{\color{black}\lstset{style=LSTSCRIBAPI}}]{@1}{@},
  moredelim=**[is][\only<2>{\color{black}\lstset{style=LSTSCRIBAPI}}]{@2}{@},
  moredelim=**[is][\only<3>{\color{black}\lstset{style=LSTSCRIBAPI}}]{@3}{@},
  moredelim=**[is][\only<4>{\color{black}\lstset{style=LSTSCRIBAPI}}]{@4}{@},
  moredelim=**[is][\only<5>{\color{black}\lstset{style=LSTSCRIBAPI}}]{@5}{@},
  moredelim=**[is][\only<6>{\color{black}\lstset{style=LSTSCRIBAPI}}]{@6}{@},
  moredelim=**[is][\only<7>{\color{black}\lstset{style=LSTSCRIBAPI}}]{@7}{@},
  moredelim=**[is][\only<8>{\color{black}\lstset{style=LSTSCRIBAPI}}]{@8}{@},
  moredelim=**[is][\only<9>{\color{black}\lstset{style=LSTSCRIBAPI}}]{@9}{@},
  moredelim=**[is][\only<1->{\color{black}\lstset{style=LSTSCRIBAPI}}]{@1-}{@},
  moredelim=**[is][\only<2->{\color{black}\lstset{style=LSTSCRIBAPI}}]{@2-}{@},
  moredelim=**[is][\only<3->{\color{black}\lstset{style=LSTSCRIBAPI}}]{@3-}{@},
  moredelim=**[is][\only<4->{\color{black}\lstset{style=LSTSCRIBAPI}}]{@4-}{@},
  moredelim=**[is][\only<1-2>{\color{black}\lstset{style=LSTSCRIBAPI}}]{@1-2}{@},  % needs to come after @1-
  moredelim=**[is][\only<2-3>{\color{black}\lstset{style=LSTSCRIBAPI}}]{@2-3}{@},
  moredelim=**[is][\only<3-4>{\color{black}\lstset{style=LSTSCRIBAPI}}]{@3-4}{@}
}

\lstdefinestyle{SCRIBAPI}%
{%
	style=SCRIBAPIBASE,
	identifierstyle=\color{black},
	stringstyle=\color{teal},
 	emphstyle=\color{dkblue}
}

\lstdefinestyle{LSTSCRIBAPI}%
{%
	style=SCRIBAPI,
	basicstyle=\LSTCODESIZE\CODESTYLE\color{black},
  %moredelim=**[is][\only<1>{\color{black!40}\lstset{style=SCRIBAPIBG}}]{@1}{@},
  moredelim=**[is][\color{black!40}\lstset{style=LSTSCRIBAPIBG}]{@1}{@},
  moredelim=**[is][\only<2>{\color{black!40}\lstset{style=LSTSCRIBAPIBG}}]{@2}{@},
  moredelim=**[is][\only<3>{\color{black!40}\lstset{style=LSTSCRIBAPIBG}}]{@3}{@},
  moredelim=**[is][\only<4>{\color{black!40}\lstset{style=LSTSCRIBAPIBG}}]{@4}{@},
  moredelim=**[is][\only<5>{\color{black!40}\lstset{style=LSTSCRIBAPIBG}}]{@5}{@},
  moredelim=**[is][\only<6>{\color{black!40}\lstset{style=LSTSCRIBAPIBG}}]{@6}{@},
  moredelim=**[is][\only<7>{\color{black!40}\lstset{style=LSTSCRIBAPIBG}}]{@7}{@},
  moredelim=**[is][\only<8>{\color{black!40}\lstset{style=LSTSCRIBAPIBG}}]{@8}{@},
  moredelim=**[is][\only<9>{\color{black!40}\lstset{style=LSTSCRIBAPIBG}}]{@9}{@},
  moredelim=**[is][\only<1->{\color{black!40}\lstset{style=LSTSCRIBAPIBG}}]{@1-}{@},
  moredelim=**[is][{\color{dkgreen}\lstset{style=ASSIST}}]{@?}{@},
}

%\lstnewenvironment{PYTHONLISTING}%
\lstdefinestyle{PYTHONAPI}%
{
%\lstset{%
	language=python,
	showstringspaces=false,
	formfeed=\newpage,
	tabsize=2,
	basicstyle=\CODESTYLE\CODESIZE,
	keywordstyle=\CODESTYLE\CODESIZE,
	%commentstyle=\color{dkgreen}\itshape\LSTFONTSIZE,
	commentstyle=\color{purple}\CODESTYLE\CODESIZE,
	stringstyle=\color{teal}\CODESTYLE\CODESIZE,
 	emphstyle=\color{dkblue}\bfseries\CODESIZE,
	morekeywords={models, lambda, forms, def, class}
	keywordstyle=\color{dkblue},
	emph={%
		access,and,as,break,class,continue,def,del,elif,else,%
		except,exec,finally,for,from,global,if,import,in,is,%
		lambda,not,or,pass,print,raise,return,try,while,assert,with%
	}
%}
}
%========================RUMI CODE MACROS==================================
\makeatletter

\newenvironment{btHighlight}[1][]
{\begingroup\tikzset{bt@Highlight@par/.style={#1}}\begin{lrbox}{\@tempboxa}}
{\end{lrbox}\bt@HL@box[bt@Highlight@par]{\@tempboxa}\endgroup}

\newcommand\btHL[1][]{%
  \begin{btHighlight}[#1]\bgroup\aftergroup\bt@HL@endenv%
}

\def\bt@HL@endenv{%
  \end{btHighlight}%
  \egroup
}
\newcommand{\bt@HL@box}[2][]{%
  \tikz[#1]{%
    \pgfpathrectangle{\pgfpoint{1pt}{0pt}}{\pgfpoint{\wd #2}{\ht #2}}%
    \pgfusepath{use as bounding box}%
    \node[anchor=base west, fill=orange!30,outer sep=0pt,inner xsep=1pt, inner ysep=0pt, rounded corners=3pt, minimum height=\ht\strutbox+1pt,#1]{\raisebox{1pt}{\strut}\strut\usebox{#2}};
  }%
}

\makeatother

\makeatletter

\newenvironment{berrHighlight}[1][]
{\begingroup\tikzset{berr@Highlight@par/.style={#1}}\begin{lrbox}{\@tempboxa}}
{\end{lrbox}\berr@HL@box[berr@Highlight@par]{\@tempboxa}\endgroup}

\newcommand\berrHL[1][]{%
  \begin{berrHighlight}[#1]\bgroup\aftergroup\berr@HL@endenv%
}

\def\berr@HL@endenv{%
  \end{berrHighlight}%
  \egroup
}

\newcommand{\berr@HL@box}[2][]{%
  \tikz[#1]{%
    \pgfpathrectangle{\pgfpoint{1pt}{0pt}}{\pgfpoint{\wd #2}{\ht #2}}%
    \pgfusepath{use as bounding box}%
    \node[anchor=base west, draw=red, fill = WhiteSmoke,,outer sep=0pt,inner xsep=1pt, inner ysep=0pt, rounded corners=3pt, minimum height=\ht\strutbox+1pt,#1]{\raisebox{1pt}{\strut}\strut\usebox{#2}};
  }%
}

\makeatother

\newcommand{\hlo}[1]{ {\sethlcolor{orange} \hl{#1}} }

\lstdefinelanguage{Scribble}{
  keywords={int},
  morecomment=[l]{//},
  morecomment=[s]{/*}{*/},
  morecomment=[s]{@"}{"},
  morestring=[b]",
  tabsize=2
}

\lstnewenvironment{SCRIBBLELISTING}%
{
\lstset{
  language=Scribble,
  xleftmargin= 2.5em,
  xrightmargin= 2.5em,
  framexleftmargin = 0.5em,
  numbers=left,
  showstringspaces=false,
  formfeed=\newpage,
  tabsize=2,
  commentstyle=\color{orange}\itshape,
  basicstyle= \SCRMATHFONT,
  keywordstyle=\color{purple}\itshape,
  emph={global, protocol, role, par, and, from, to, choice, at, or, rec, continue},
  morekeywords = [2]{A,B,C,D,E, S, P},
  keywordstyle = [2]\color{beige}\itshape,
  emphstyle=\color{dkblue}\bfseries,
  basicstyle=\CODESTYLE\scriptsize,
  %frame=single,
  escapeinside={*@}{@*},
  moredelim=**[is][\btHL]{`}{`},
}
}
{
}

\lstdefinestyle{myocaml}{
  style=BASE,
  keepspaces,
  language=SOCaml,
  tabsize=2,
  numbersep=1mm,
  keywordstyle = \color{dkblue},
 % numbers=left,
  basicstyle=\CODESIZETINY\CODESTYLE,
  morekeywords = [3]{choice_at, finish},
  keywordstyle = [3]\color{purple},
  keywordstyle = [4]\color{dkred},
  %basicstyle=\CODESIZE, %\CODESTYLE\fontsize{9}{7}\selectfont,
  %basicstyle=\CODESIZE, %\CODESTYLE\fontsize{7}{8}\selectfont,
  commentstyle = \color{dkgreen}\itshape,
  alsoletter = {:,->,(,), _ },
  escapeinside={^}{^},
  mathescape = true,
  literate=
    {-->}{{\color{purple}-{}->}}1
    %% {-->}{$\color{purple}{\longrightarrow}$}1
    {:?}{$\color{blue}{:? \quad}$}1
    {<}{<}1
%    {\%}{{\color{dkred}\%}}1
     {@@}{{\color{dkblue}@@}}1
    {<@}{{\color{blue}<@} \quad}1
    {@>}{\quad{\color{blue}@}>}1
    {>}{>}1
    {._}{{$\,$}}1
    {,,}{{$\,\elipc\,$}}1
    {**}{{,}}1
    {???}{{\%}}1
    %{_1}{$_\textsf{\scriptsize 1}$}1
}

\lstnewenvironment{OCAMLLISTING}%
{\lstset{style=myocaml}}
{}

\lstnewenvironment{OCAMLLISTINGTINY}%
{\lstset{style=myocaml,basicstyle=\scriptsize\CODESTYLE}}
{}

\lstnewenvironment{BIGOCAMLLISTING}%
{\lstset{style=tCODEstyle}}
{}

\lstdefinestyle{tCODEstyle}{
    style=BASE,
    language = SOCaml,
    alsoletter = {_},
    alsoletter = {'},
    basicstyle=\CODESIZE\CODESTYLE,
    emph={string,str,int, real, bool, unit},
    morekeywords = [2]{dynlin, lin, data},
    keywordstyle = [2]\color{gray},
    emphstyle=\color{dkred},
    keywordstyle=\CODESIZE\CODESTYLE\color{dkred},
    mathescape=true,
    breaklines=true,
    %backgroundcolor=\color{blue},
    literate=
    {-->}{{\color{purple}-{}->}}1
    %% {-->}{$\color{purple}{\longrightarrow}$}1
    {:?}{$\color{dkblue}{:? }$}1
    {<}{<}1
     {(}{(}1
    {\%}{{\color{blue}\%}}1
     {@@}{{\color{dkblue}@@}}1
    {<@}{{\color{blue}<@} \quad}1
    {@>}{\quad{\color{blue}@}>}1
    {>}{>}1
    {._}{{$\,$}}1
   % {_1}{$_\textsf{\scriptsize 1}$}1
}

\definecolor{mygray}{RGB}{245,245,245}
\lstnewenvironment{OCAMLLISTINGDARK}%
{
\lstset{
  language=SOCaml,
  %backgroundcolor = \color{mygray},
  alsoletter ={_},
  alsoletter ={.},
  tabsize=2,
  numbersep=1mm,
  keywordstyle = \color{dkblue},
 % numbers=left,
  basicstyle=\tiny\CODESTYLE,
  morekeywords = [3]{choice_at, finish},
  keywordstyle = [3]\color{purple},
  keywordstyle = [4]\color{dkred},
  %basicstyle=\CODESIZE, %\CODESTYLE\fontsize{9}{7}\selectfont,
  %basicstyle=\CODESIZE, %\CODESTYLE\fontsize{7}{8}\selectfont,
  commentstyle = \color{dkgreen}\itshape,
  alsoletter = {:,->,.,(,)},
  %morecomment=[1]{//},
  escapeinside={^}{^},
  mathescape = true,
  %moredelim=**[is][\berrHL]{`}{`},
  moredelim=**[is][\btHL]{~}{~},
  literate=
    {-->}{{\color{purple}-{}->}}1
    %% {-->}{$\color{purple}{\longrightarrow}$}1
    {:?}{$\color{blue}{:? \quad}$}1
    {<}{<}1
     {(}{(}1
    {\%}{{\color{blue}\%}}1
     {@@}{{\color{dkblue}@@}}1
    {<@}{{\color{blue}<@} \quad}1
    {@>}{\quad{\color{blue}@}>}1
    {>}{>}1
    {._}{{$\,$}}1
    {,,}{{$\,\elipc\,$}}1
    {**}{{,}}1
    {_1}{$_\textsf{\scriptsize 1}$}1
}
}
{
}

\lstdefinestyle{oCODEstyle}{
    style=BASE,
    keepspaces,
    language = SOCaml,
    alsoletter = {_},
    basicstyle=\CODESIZENORMAL\CODESTYLE,
    emph={string,str,int, real, bool, unit},
    emphstyle=\color{dkred},
    keywordstyle=\CODESIZENORMAL\CODESTYLE\color{dkblue},
    mathescape=true,
    breaklines=true,
    %backgroundcolor=\color{blue},
    literate=
    {-->}{{\color{purple}-{}->}}1
    %% {-->}{$\color{purple}{\longrightarrow}$}1
    {:?}{$\color{dkblue}{:? }$}1
    {<}{<}1
     {(}{(}1
    {\%}{{\color{blue}\%}}1
    {@@}{{\color{dkblue}@@}}1
    {<@}{{\color{blue}<@} \quad}1
    {@>}{\quad{\color{blue}@}>}1
    {>}{>}1
    {._}{{$\,$}}1
    {,,}{{$\,\elipc\,$}}1
    {**}{{,}}1
   % {_1}{$_\textsf{\scriptsize 1}$}1
}
\newcommand{\oCODEEsc}[1]{\lstinline[style=oCODEstyle]!#1!}
\newcommand{\oCODE}{\lstinline[style=oCODEstyle]}

\newcommand{\sCODE}{\lstinline[
	basicstyle=\CODESIZE\CODESTYLE,
  keywords={from,to, interrupt, by, choice, at, and,
  rec, continue, global, protocol, local, role, par, introduces, as, interruptible},
  	emph={string,str,int, real, bool},
  	emphstyle=\color{red},
  keywordstyle=\CODESIZE\CODESTYLE\color{blue},
  mathescape=true,
  breaklines=true,
  backgroundcolor=\color{blue}
]}

\mdfdefinestyle{myframedcodestyle}{
  skipabove=4pt,
  skipbelow=2pt,
  leftmargin=0pt,
  rightmargin=0pt,%
  innertopmargin=0pt,%
  innerbottommargin=0pt,
  innerleftmargin=-1em,
  innerrightmargin=0pt,
  %% ignorelastdescenders,%
  %% settings={\lstset{resetmargins}},%
}
\newenvironment{myframedcode}[1][]
  {\begin{mdframed}[style=myframedcodestyle,#1]}
  {\end{mdframed}}

\lstdefinestyle{myocamltight}{
    style=myocaml,
    xleftmargin=2em,
    xrightmargin=2em,
    belowskip=2pt,
    aboveskip=4pt,
    framexleftmargin=0pt,
    framexrightmargin=0pt,
    framextopmargin=0pt,
    framexbottommargin=0pt,
    resetmargins=true,
  }

\lstnewenvironment{LISTING}[1][]%
{
  \lstset{
    style=myocamltight,
    xleftmargin=1em,
    xrightmargin=1em,
    belowskip=0pt,
    #1
  }
}{}

\lstnewenvironment{LISTINGNORMAL}[1][]%
{
  \lstset{
    style=myocamltight,
    xleftmargin=1em,
    xrightmargin=1em,
    belowskip=0pt,
    basicstyle=\CODESIZENORMAL\CODESTYLE,
    keywordstyle=\CODESIZENORMAL\CODESTYLE\color{dkblue},
    #1
  }
}{}

\lstnewenvironment{LISTINGBOX}[1][]%
{
  \lstset{
    style=myocamltight,
    frame=single,
    #1
  }
}{}

% Scribble grammar
%\newcommand{\ptp}[1]{\mathtt{\color{blue}#1\color{black}}}
%\newcommand{\key}[1]{\mathtt{\color{dkblue}#1\color{black}}}
%\newcommand{\trm}[1]{\mathtt{\color{red}#1\color{black}}}
%\newcommand{\SORT}{\mathtt{S}}
%\newcommand{\mesg}[2]{\mathtt{#1}(\mathtt{#2})}
%
%\newcommand{\info}[1]{{\newline \textbf{Additional info:} #1}}

%% \renewcommand{\lstlistingname}{List.}
%% % Fix counter as described at https://tex.stackexchange.com/a/28334/9075
%% %\usepackage{chngcntr}
%% \AtBeginDocument{\counterwithout{lstlisting}{section}}
\newcommand{\mrg}{\mathit{mrg}}

\section{Introduction}
Multiparty session types (MPST)~\cite{HondaYC08} are a popular
type-driven technique to ensure the correctness of concurrent programs
that communicate using message-passing.
The key benefit of MPST is to guarantee statically that the components of
a program have compatible behaviours,
%abides by an agreed communication protocol
and thus no components can get permanently stuck.
Many implementations of MPST in different programming languages have
been proposed in the last
decade~\cite{NgCY15,HuY16,KouzapasDPG16,ScalasDHY17,NHYA2018,CastroHJNY19,ImaiNYY19,miuGenerating2020,harveyMultiparty2021,DBLP:journals/pacmpl/00020HNY20},
however, all suffer from a notable shortcoming: they require
programmers to adopt a top-down approach that does not fit well in
modern development practices.
%
% Indeed 
When changes are frequent and continual (e.g., continuous delivery),
re-designing the program and its specification at every change is not
feasible.

Most MPST theories and tools advocate an intrinsically top-down
approach.
They require programmers to specify the communication (often in the
form of a global type) of their programs before they can be
type-checked.
In practice, type-checking programs against session types is very
difficult.
To circumvent the problem, most implementations of MPST rely on
\emph{external} toolings that generate code from a global type, see e.g.,
all works based on the Scribble toolchain~\cite{YHNN2013}.
%
% This is undesirable for two reasons:
% %
% (1) the lan global types
% 
% are far from the types of a “mainstream” programming
% language, state-of-the-art MPST implementations use external
% domain-specific protocol description languages and tools (e.g. the
% Scribble toolchain \cite{YHNN2013}) to specify global types and to
% generate protocol-specific communication APIs for programming
% message-passing programs.
%

% be expressed as a global type,
% and hence cannot be implemented with the existing MPST libraries. 
% Second, as shown in~\cite{KMC}, global types have limited expressivity,
% many otherwise correct concurrency programs cannot be expressed as a global type,
% and hence cannot be implemented with the existing MPST libraries.

% Top-down development does not fit well in modern development
% practice where changes are frequent and continual (e.g., continuous
% delivery); and thus re-designing the program and its specification at
% every change is not feasible.

In this paper, we present an OCaml library, called {\tt kmclib}, that
supports the development of programs which enjoy all the benefits of
MPST while avoiding their main drawbacks.
The {\tt kmclib} library guarantees that threads in well-typed programs
will not get stuck. The library also enables \emph{bottom-up development}:
programmers write message-passing programs in a natural way, without
having to write session types.
Our library is built on top of {\em Multicore
  OCaml}~\cite{DBLP:journals/pacmpl/Sivaramakrishnan20} that 
offers highly scalable and efficient concurrent programming, but
%% , including asynchronous I/O via
%% {\em fibers} (lightweight threads akin to coroutines) \cite{DBLP:conf/pldi/Sivaramakrishnan21}.
% \KInote{Omitted fibers; as they (eff. handlers) are a separate work from Multicore OCaml and irrelevant to this paper.}
% However, Multicore
% OCaml
does not provide any static guarantees wrt.\ concurrency.

% However, both lacks support for static checking on concurrency, in
% spite of its extensive tradition on types.

% OCaml nowadays is a venue for modern highly scalable concurrent
% programming with static typing, on top of which \kmclib{} is built.
% 
%% its known safety backed by the type system with (extended) ML-style polymoprhism,
%% an industrial-strength programming language with its tradition of (single-threaded) runtime efficiency,
%% and used for not only compiler development and verification toolchain
%% and used for system programming in both
%% industry and academia ~\cite{mldonkey,madhavapeddy08xen,minsky11ocaml,madhavapeddy14unikernels,radanne16eliom}.
% Recently, {\em Multicore OCaml}~\cite{DBLP:journals/pacmpl/Sivaramakrishnan20} is integrated into
% OCaml with a support for multiple parallel thread based on its
% efficient concurrent GC.  The {\em effect handler} extension
% \cite{DBLP:conf/pldi/Sivaramakrishnan21} offers efficient asynchronous
% I/O with {\em fibers} (lightweight threads akin to coroutines).
% However, both lacks support for static checking on concurrency, in
% spite of its extensive tradition on types.

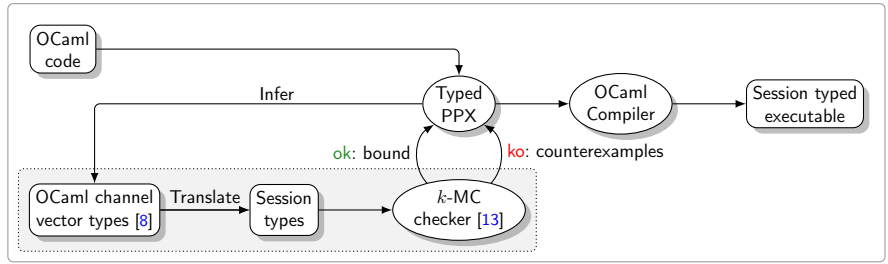
\begin{figure}[t!]
  \centering
  \begin{tikzpicture}[scale=0.7, every node/.style={transform shape},node distance=0.9cm and 1.4cm,tips=proper]
    \node[deliv] (code) {OCaml \\ code}; 
    \node[manip, below right=of code,xshift=5cm,yshift=0.7cm] (ppx) {Typed \\ PPX};
    \node[manip, right=of ppx,xshift=0cm] (compile) {OCaml \\ Compiler};
    \node[deliv, right=of compile] (exec) {Session typed \\ executable};
    \node[manip, below=of ppx] (kmc) {\kmc \\ checker{\small ~\cite{LangeY19}}};
    \node[deliv, left=of kmc] (stypes) {Session \\ types};    
    \node[deliv, left=of stypes,xshift=-0.3cm] (chan) {OCaml channel \\ vector types{\small ~\cite{ImaiNYY19}}};

    % \node[deliv, right=of kmc] (res) {Bound\\ or trace};
    %
    % \node[above of=chan] (tmp) {t};
    \draw[line] (code) -| (ppx);
    \draw[line] (ppx) -- (compile);
    \draw[line] (compile) -- (exec);
    \draw[line] (chan) -- (stypes);
    \draw[line] (ppx) -| (chan);
    \node at  (ppx-|chan) (txt) {};
    \node[yshift=0.2cm] (lab) at ($(txt)!0.5!(ppx)$) {Infer};
    % 
    % \draw[line]
    % (ppx.south west) edge node [above] {Infer} (chan);
    % % % 
    \path[line]
    (chan) edge node [above] {Translate} (stypes);
    \path[line]
    (stypes) edge node [above] {} (kmc);
    \path[line]
    (kmc) edge [bend left=50] node [left] {{\color{ForestGreen} ok}: bound} (ppx);
    \path[line]
    (kmc) edge [bend right=50] node [right] {{\color{red} ko}: counterexamples} (ppx);   
    \begin{scope}[on background layer]
      \node[draw=black, densely dotted, rounded corners=2pt,
      fit=(chan) (kmc),inner sep=4pt,fill=gray!10] (incheck) {};
    \end{scope}
    \begin{scope}[on background layer]
      \node[draw=black!40, rounded corners=2pt,
      fit=(code) (exec) (chan) (kmc),inner sep=8pt] (outcheck) {};
    \end{scope}
  \end{tikzpicture}
  \caption{Workflow of the \kmclib library.}
  \label{fig:workflow}
\end{figure}

Figure~\ref{fig:workflow} gives an overview of \kmclib.
Its implementation combines the power of the {\em type-aware} macro
system of OCaml (Typed PPX) with two recent advances in the session types area:
%
% a state-of-the-art implementation 
an encoding of MPST in OCaml (channel vector types~\cite{ImaiNYY19}) 
and a session type compatibility
checker (\kmc checker~\cite{LangeY19}).
To our knowledge, this is the first implementation of type inference
for MPST and the first integration of compatibility
checking in a programming language.

\newcommand{\filepic}{
  \begin{tikzpicture}
    \foreach \i in {1,2,3} {
      \begin{scope}[shift={(.2*\i,.2*\i)}]
        \draw[bottom color=black!7, top color=white, drop shadow={shadow xshift=-.4ex}]
        (0,0) -- ++(3,0) -- ++(0,3)  -- ++(-1,1) -- ++(-2,0) -- cycle;
        \draw (3,3) -| (2,4);
      \end{scope}
    }
  \end{tikzpicture}
}
  
%%% Local Variables:
%%% mode: latex
%%% TeX-master: "main"
%%% End:

% \JLnote{
% Additionally not all communication patterns fit the synchrony-oriented
% syntax of global types, i.e., many correct message-passing programs
% cannot be represented as a global type (e.g., fire-and-forget
% patterns~\cite{LangeY19}).}

The {\tt kmclib} library offers several advantages compared to earlier MPST implementations.
\textbf{(1)} It is \emph{flexible}: 
% programs do not have to implement
% communication patterns that fit the synchrony-oriented syntax of
% global types.
programmers can implement communication patterns (e.g.,
fire-and-forget patterns~\cite{LangeY19}) that are not expressible in
the synchrony-oriented syntax of global types.
%
% As it does not rely on global types, it enables the implementation of programs that 
% are manifold: it is (1) expressive --
% it does not rely on specifying a global type, but utilises a state-of-the-art
% compatibility checker, as such it can implement and verify strictly more programs than any other
% MPST implementations;
%
\textbf{(2)} It is \emph{lightweight} as it piggybacks on OCaml's type
system to check and infer session types, hence lifting the burden of
writing session types off the programmers.
\textbf{(3)} It is \emph{user-friendly} thanks to its integration in
Visual Studio Code, e.g., compatibility violations are mapped to
precise locations in the code.
\textbf{(4)} It is \emph{well-integrated} into the natural
edit-compile-run cycle. Although compatibility is checked by an
external tool, this step is embedded as a compilation step and thus
hidden from the user.

% thus developers themselves do not
% have to use any external tools to aid the verification process.
 
% \rutodo{I moved the workflow pic so we should say smth about it here or in the beginning of section 2}
% Benefits:
% \begin{itemize}
% \item more expressive than global type (MOST expressive)
% \item little to no overhead for programmer (BEST integration)
% \item good error messages / integration with IDE (VSCode)
% \end{itemize}
 
% In this paper we present the first tool that combines both approaches. Explain the tool in a few sentences here\dots
% \textbf{Contributions}
% \begin{itemize}
% \item library for safe concurrent programming in OCaml
% \item extension to multi-core OCaml
% \item efficient (system) programming in ML with a static guarantee on concurrency.
% \item deadlock-freedom by-construction, or deadlock-freedom by inference.
% \item buffer utilisation (?)
% \item new typed ppx extension
% \end{itemize}
% Plan:
% \begin{itemize}
% \item explain usage of tool with example
% \item explain how it works afterwards (ppx then kmc)
% \item do some related work
% \end{itemize}
 
%%% Local Variables:
%%% mode: latex
%%% TeX-master: "main"
%%% End:

\section{Safe Concurrent Programming in Multicore OCaml}
%% This section explain the library by example.
%% The example is give in Fig. X.

We give an overview of the features and usage of \kmclib using the
program in Figure~\ref{fig:fibworker} (top) which calculates Fibonacci numbers.
The program consists of three concurrent \emph{threads}
(\texttt{user}, \texttt{master}, and \texttt{worker}) that interact
using point-to-point message-passing.
Initially, the \texttt{user} thread sends a request to the
\texttt{master} to start the calculation, then waits
for the \texttt{master} to return a work-in-progress message, or
the final result. After receiving the result, the \texttt{user} sends
back a stop message.
Upon receiving a new request, the \texttt{master} splits the initial
computation in two and sends two tasks to a
\texttt{worker}. For each task that the \texttt{worker} receives, it
replies with a result.  The \texttt{master} and \texttt{worker}
threads are recursive and terminate only upon receiving a stop
message.

Figure~\ref{fig:fibworker} (bottom) gives a session type for each
thread, i.e., the behaviour of each thread wrt.\ communication.
For clarity we represent session types as a communicating finite state
machine (CFSM~\cite{cfsm83}), where !  (resp.\ ?) denotes sending
(resp.\ receiving).
For example, $\PSEND{um}{compute}$ means that the user is sending to
the master a message \msg{compute}, while $\PRECEIVE{um}{compute}$
says that the master receives \msg{compute} from the user.
Our library infers these CFSM representations from the OCaml code, in
Figure~\ref{fig:fibworker} (top), and verifies \emph{statically} that
the three threads are \emph{compatible}, hence no thread can get stuck
due to communication errors.
If compatibility cannot be guaranteed, the compiler reports the kind
of violations (i.e., \emph{progress} or \emph{eventual reception}
error) and their locations in the code.
Figure~\ref{fig:error-progress} shows how such semantic errors are
reported visually in Visual Studio Code.

Albeit simple, the common communication pattern used in
Figure~\ref{fig:fibworker} cannot be expressed as a global type, and
thus cannot be implemented in previous MPST implementations.
Concretely, global types cannot express the intrinsic asynchronous
interactions between the master and worker threads (i.e., the master
may send a second task message, while the worker sends a result).
 
% just a copy-paste from https://github.com/keigoi/jrklib/blob/master/test/paper/test.ml

\begin{figure}[t]
 \begin{minipage}{0.01\textwidth}
 \end{minipage}
\begin{minipage}{0.49\textwidth}
\begin{LISTING}[basicstyle=\scriptsize\CODESTYLE,numbers=left]
let KMC (uch,mch,wch) = [--?%?--kmc.gen (u,m,w)]^\label{line:kmcgen}^

let user () = ^\label{line:userbegin}^
  let uch = send uch#m#compute 42 in ^\label{line:usersend}^
  let rec loop uch : unit =
    match receive uch#m with ^\label{line:usrrcv}^
    | `wip(res, uch) -> ^\label{line:usermatchbegin}^
      printf "in progress: %d\n" res;
      loop uch
    | `result(res, uch) -> ^\label{line:usermatchend}^
      printf "result: %d\n" res;
      send uch#m#stop () ^\label{line:usrrcvend}^
  in loop uch ^\label{line:userend}^

let worker () =^\label{line:workerbegin}^
  let rec loop wch : unit =
    match receive wch#m with ^\label{line:workermatchbegin}^
    | `task(num, wch) ->
      loop (send wch#m#result (fib num))
    | `stop((), wch) -> wch^\label{line:workermatchend}^
  in loop wch^\label{line:workerend}^   ^\ \newpage ^
\end{LISTING}
\end{minipage}
\begin{minipage}{0.46\textwidth}
\begin{LISTING}[basicstyle=\scriptsize\CODESTYLE,numbers=left,firstnumber=last]
let master () =^\label{line:masterbegin}^
  let rec loop (mch --!: [%kmc.check u]!--) : unit =^\label{line:masterloopstart}^
    match receive mch#u with
    | `compute(x, mch) ->
      let mch = send mch#w#task (x - 2) in ^\label{line:commented-progress}^
      let mch = send mch#w#task (x - 1) in
      let `result(r1, mch) = receive mch#w in
      let mch = send mch#u#wip r1 in
      let `result(r2, mch) = receive mch#w in ^\label{line:stuck-progress}^
      loop (send mch#u#result (r1 + r2)) ^\label{line:evt-recept-viol}^
    | `stop((), mch) ->
      send mch#w#stop ()
  in loop mch^\label{line:masterend}^

let () = ^\label{line:threadstart}^
  let ut = Thread.create user () in
  let mt = Thread.create master () in
  let wt = Thread.create worker () in  ^\label{line:threadend}^
  List.iter Thread.join [ut;mt;wt] ^\label{line:join}^ 
\end{LISTING}
\end{minipage}
\vspace{2ex}

\hrule

% \begin{figure}[t]
%   \centering
  \begin{tabular}{c@{\;}c@{\;}c}
    ${\large \ptp{u}}$:
    \begin{tikzpicture}[mycfsm, node distance=0.5cm and 0.5cm]
      \node[state, initial, initial where=above] (s0) {};
      \node[state, below=of s0] (s1) {};
      \node[state, below=of s1] (s2) {};
      \node[state, right=of s2] (s3) {};
      \path
      (s0) edge node [left] {$\PSEND{um}{compute}$} (s1)
      (s1) edge node [left] {$\PRECEIVE{mu}{result}$} (s2)
      (s2) edge node [below] {$\PSEND{um}{stop}$} (s3)
      (s1) edge [loop right] node [right] {$\PRECEIVE{mu}{wip}$} (s1)
      ;
    \end{tikzpicture}
    &
      ${\large \ptp{m}}$:
      \begin{tikzpicture}[mycfsm, node distance=0.5cm and 0.9cm]
        \node[state, initial, initial where=above] (s0) {};
        \node[state, below=of s0] (s8) {};
        \node[state, below left=of s8] (s1) {};
        
        \node[state, left=of s1] (s2) {};
        \node[state, left=of s2] (s3) {};
        \node[state, above=of s3] (s4) {};
        \node[state] at  (s3|-s0) (s5) {};
        \node[state, below right=of s0] (s6) {};
        \node[state, below=of s6] (s7) {};
        \path
        (s0) edge node [left] {$\PRECEIVE{um}{compute}$} (s8)
        (s8) edge node [below,sloped] {$\PSEND{mw}{task}$} (s1)
        (s1) edge node [] {$\PSEND{mw}{task}$} (s2)
        (s2) edge node [] {$\PRECEIVE{wm}{result}$} (s3)
        (s3) edge node [right] {$\PSEND{mu}{wip}$} (s4)
        (s4) edge node [right] {$\PRECEIVE{wm}{result}$} (s5)
        (s5) edge node [] {$\PSEND{mu}{result}$} (s0)
        (s0) edge node [sloped] {$\PRECEIVE{um}{stop}$} (s6)
        (s6) edge node [] {$\PSEND{mw}{stop}$} (s7)
        ;
      \end{tikzpicture}
    &
      ${\large \ptp{w}}$:
      \begin{tikzpicture}[mycfsm, node distance=0.5cm and 0.7cm]
        \node[state, initial, initial where=above] (s0) {};
        \node[state, below left=of s0] (s1) {};
        \node[state, below right=of s0] (s2) {};
        \path
        (s0) edge [bend right=90] node [above,yshift=0.1cm] {$\PRECEIVE{mw}{task}$} (s1)
        (s0) edge [bend left=90] node [above,sloped]  {$\PRECEIVE{mw}{stop}$} (s2)
        (s1) edge [bend right=90] node [below]  {$\PSEND{wm}{result}$} (s0)
        ;
      \end{tikzpicture}
  \end{tabular}
  % \caption{Session types inferred from Fig.~\ref{fig:fibworker}.} 
%   \label{fig:cfsms}
% \end{figure}

%%% Local Variables:
%%% mode: latex
%%% TeX-master: "main"
%%% End:

  \caption{Example of {\tt kmclib} program (top) and \emph{inferred} session types (bottom).}\label{fig:fibworker}
  
\end{figure}

%%% Local Variables:
%%% mode: latex
%%% TeX-master: "main"
%%% End:

\myparagraph{Programming with \kmclib}
% Figure~\ref{fig:fibworker} shows the implementations of the three threads using \kmclib.
To enable safe message-passing programs, \kmclib provides two
communication primitives, \oCODE{send} and \oCODE{receive}, and two
primitives for channel creation (\oCODE{KMC} and \oCODE{%kmc.gen}).
  We only give a user-oriented description of these primitives here
  (see Appendix~\ref{sec:appimpl} an overview of their implementations).

  % primitives and their underlying representation.

  % explain them by example for the general programmer, omitting
  % OCaml-specific details.
  % % 
  % The OCaml-savvy programmers may refer to \S~\ref{sec:impl} for
  % technical explanation of the primitives and their underlying
  % representation.
  % 
  % Each primitive expects a communication channel, the name of the destination process (for sending/receiving),
  % a message label and a payload. Having labels is
  % needed as to distinguish multiple possible receives.
  
  The crux of \kmclib is the {\bf session channel creation}:
  \oCODE{[%kmc.gen (u,m,w)]}
    at Line \ref{line:kmcgen}.
    This primitive takes a tuple of \emph{role names} as argument
    (i.e., \oCODE{(u,m,w)}) and returns a tuple of communication
    channels, which are bound to \oCODE{(uch,mch,wch)}.
    These channels will be used by the threads implementing roles
    \texttt{user} (Lines~\ref{line:userbegin}-\ref{line:userend}),
    \texttt{worker} (Lines~\ref{line:workerbegin}-\ref{line:workerend}),
    and
    \texttt{master} (Lines~\ref{line:masterbegin}-\ref{line:masterend}).
    By default, channels are implemented using concurrent queues from
    Multicore OCaml (\oCODE{Domainslib.Chan.t}) but other underlying
    transports can easily be provided.
    % 

    % JL: NOT HERE :)
    % \oCODE{KMC} is a type constructor for the communication channels.
    % \footnote{The \oCODE{KMC} constructor is required for the inferred type to be {\em monomorphic}. See \S~\ref{subsec:ppx}.}

    Threads send and receive messages over these channels using the
    communication primitives provided by \kmclib. The send primitive
    requires three \emph{arguments}: a channel, a destination role,
    and a message.
    For instance, the \texttt{user} sends a request to the
    \texttt{master} with \oCODE{send uch#m#compute 20} where
    \oCODE{uch} is the user's communication channel, \oCODE{m} indicates the
    destination, and \oCODE{compute 20} is the message (consisting
    of a label and a payload).
    Observe that a sending operation returns a new channel which is to
    be used in the continuation of the interactions, e.g., \oCODE{uch}
    bound at Line~\ref{line:usersend}.
    Receiving messages work in a similar way to sending messages,
    e.g., see Line~\ref{line:usrrcv} where the~\texttt{user} waits for
    a message from the~\texttt{master} with \oCODE{receive uch#m}.
    %
    % For example, the~\texttt{user}  receives from the~\texttt{master}
    % with \oCODE{receive uch#m}.
    % 
    % Note that more than one message can be received.
    We use OCaml's pattern matching to match messages against their
    labels and bind the payload and continuation channel.
    See, e.g.,
    Lines~\ref{line:usermatchbegin}-\ref{line:usermatchend} where
    the~\texttt{user} expects either \oCODE{wip} or \oCODE{result}
    message.
    The receive primitive returns the payload \oCODE{res} and a new
    communication channel \oCODE{uch}.
    % 
    %% See, e.g.,
    %% Lines~\ref{line:workermatchbegin}-\ref{line:workermatchend} where
    %% the worker thread expects either a new \oCODE{task} or a
    %% \oCODE{stop} signal.
    
    New thread instances are spawned in the usual way; see
    Lines~\ref{line:threadstart}-\ref{line:threadend}.
    The code at Line~\ref{line:join} waits for them to terminate.
    
    % \begin{wrapfigure}{r}{0.53\textwidth}
    \begin{figure}[t!]
      \includegraphics[width=6cm]{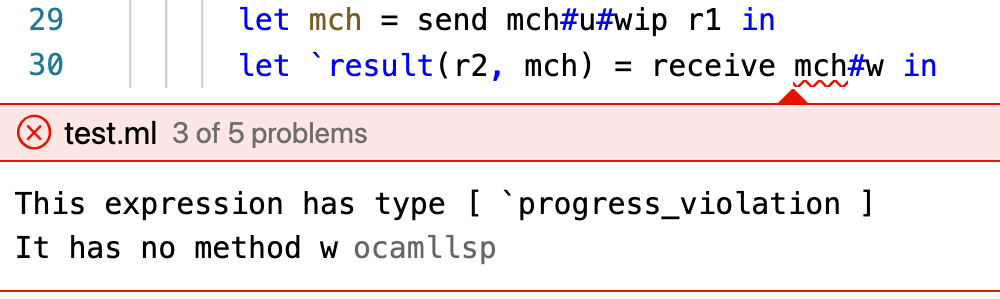}
      \;
      \includegraphics[width=6cm]{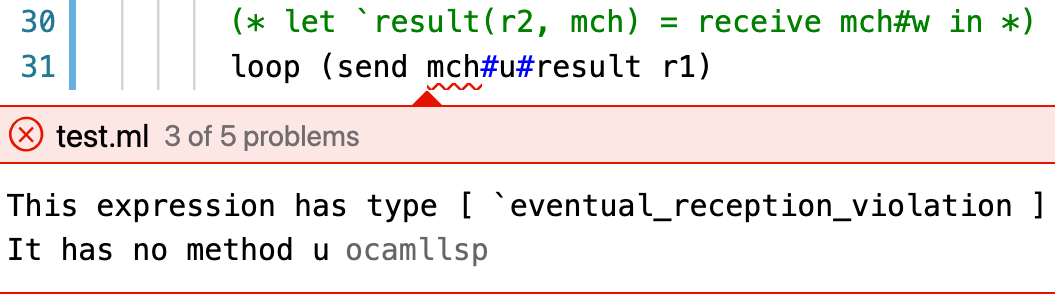}
      \caption{Examples of type errors.}\label{fig:error-progress}
    \end{figure}
    % \end{wrapfigure}
    
    \myparagraph{Compatibility and error reporting}
    While the code in Figure~\ref{fig:fibworker} may appear unremarkable,
    it hides a substantial machinery that guarantees that, if a program
    type-checks, then its constituent threads are safe, i.e., no thread
    gets permanently stuck and all messages that are sent are eventually
    received.
    This property is ensured by \kmclib using OCaml's type inference
    and PPX plugins to infer a session type from each thread then check
    whether these session types are $k$-\emph{multiparty compatible}
    (\kmc)~\cite{LangeY19}.

    % \textbf{Step~(1):} infer a session type from each thread.  The
    % library relies on the {\em type} of \oCODE{[%kmc.gen]}
    % (Line~\ref{line:kmcgen}) which reflects {\em all} behavioural
    % information of the system via OCaml's type inference.
    %   % 
    % The entire {\em usage} of variables \oCODE{uch}, \oCODE{mch} and
    % \oCODE{wch} in Lines \ref{line:userbegin}-\ref{line:masterend} is
    % gathered during type-checking.
    % 
    If a system of session types is \kmc, then it is
    safe~\cite[Theorem~1]{LangeY19}, i.e., it has the \emph{progress}
    property (no role gets permanently stuck in a receiving state) and
    the \emph{eventual reception} property (all sent messages are
    eventually received).
    Checking \kmc notably involves checking that all their executions (where
    each channel contains at most $k$ messages) satisfy progress and
    eventual reception.
    
    The \kmc-checker~\cite{LangeY19} performs a bounded verification to discover the
    \emph{least} $k$ for which a system is $k$-MC, up-to a specified upper
    bound $N$.
    In the \kmclib API, this bound can be optionally specified with
    %\rutodo{Mention Bounds here?}
    % Optionally, the channel bounds (maximum length of the queue) $k$
    %  can be specified as
    \oCODE{[%kmclib.gen $\mathit{roles}$ ~bound:$N$]}.
      The \kmc-checker emits an error if the bound is insufficient to
      guarantee safety.
      % i.e a senders may get stuck.

    % 
    % If the verification fails, it returns counter-example traces that
    % violate progress or eventual reception.

      The \oCODE{[%kmc.gen (u,m,w)]}
        primitive also feeds the results of \kmc checking back to the
        code.
      If the inferred session types are \kmc, then channels for roles
      \oCODE{u}, \oCODE{m} and \oCODE{w} can be generated.
      If \kmc cannot be guaranteed, then this results in a type error.
      We have modified the \kmc-checker to return counterexample traces
      when the verification fails.
      This helps give actionable feedback to the programmer, as
      counterexample traces are translated to OCaml types and inserted
      at the hole corresponding to \oCODE{[%kmc.gen]}.
        This has the effect of reporting the precise location of the
        errors.
        
        To report errors in a function parameter, we provide an
        \emph{optional} macro for types: \oCODE{[%kmc.check rolename]}
          (see faded code in Line~\ref{line:masterloopstart}).
        % \footnote{To report the errors in a function parameter,
        % we provide additional macro for {\em types}, \oCODE{[\%kmc.check rolename]},
        % which is placed at Line~\ref{line:masterloopstart} with grey letter.
        % }.
        %
        Figure~\ref{fig:error-progress} shows examples of such error
        reports.
        The left-hand-side shows the reported error when
        Line~\ref{line:commented-progress} is commented out, i.e., the
        master sends one task, but expects two result messages; hence
        progress is violated since the master gets stuck at
        Line~\ref{line:stuck-progress}.
        The right-hand-side shows the reported error when
        Line~\ref{line:stuck-progress} is commented out.
        In this case, variable \oCODE{mch} in
        Line~\ref{line:evt-recept-viol} (\texttt{master}) is
        highlighted because the \texttt{master} fails to consume a
        message from channel \oCODE{mch}.  

\section{Inference of Session Types in  \kmclib}
\label{sec:impl}
% We give a detailed account of the design and implementation of our library.
%Two insihghts underpin the development of \kmclib -- 
%meta-programming facilities and structural typing.  
%
%Figure~\ref{fig:workflow} gives an overview of its architecture.
%

% \begin{itemize}
% \item structurally typed object
% \item Row polymorphism
% \item  PPX
% \end{itemize}

% \subsection{The \kmclib{} API}\label{subsec:kmclib}
%
\myparagraph{The \kmclib{} API} The \kmclib{} primitives allow the
vanilla OCaml typechecker to infer the session structure of a program,
while simultaneously providing a user-friendly communication API for the
programmer.
To enable inference of session types from concurrent programs,
we leverage OCaml's structural typing and row polymorphism.
In particular, we reuse the encoding from~\cite{ImaiNYY19} where input
and output session types are encoded as polymorphic variants and
objects in OCaml.
In contrast to~\cite{ImaiNYY19} which relies on programmers writing
global types prior to type-checking, \kmclib{} infers and verifies
local session types automatically, without requiring any additional
type or annotation.

\myparagraph{Typed PPX Rewriter}
To extract and verify session types from a piece of OCaml code, the
\kmclib library makes use of OCaml PreProcessor eXtensions (PPX)
plugins which provide a powerful meta-programming facility.
PPX plugins are invoked during the compilation process to manipulate
or translate the abstract syntax tree (AST) of the program. This is
often used to insert additional definitions, e.g., pretty-printers, at
compile-time.

A key novelty of \kmclib is the combination of PPX with a form
of {\em type-aware translation}, whereas most PPX plugins
typically perform purely syntactic (type-unaware) translations.
%
% This part of the library is implemented in a new PPX plugin.
% Its behaviour is
% as follows.
%
Figure~\ref{fig:inference} shows the workflow of the PPX rewriter,
overlayed on code snippets from Figure~\ref{fig:fibworker}. The
inference works as follows.
%
%
% The verification pipeline of the {\tt kmclib} PPX plugin 
% is shown in Figure~\ref{fig:inference}, 
% (partly overlayed on the program) and it is explained below. 
\begin{enumerate}
\item The plugin reads the AST of the program code to replace the
  \oCODE{[%kmc.gen]}
    primitive with a {\em hole}, which can have {\em any} type.
  \item The plugin invokes the {\em typechecker} to get the {\em
      typed} AST of the program.
    In this way, the type of the hole is {\em inferred} to be a tuple
    of {\em channel object types} whose structure is derived
    from their usages (i.e., \oCODE{mch#u#compute}).

    To enable this propagation, we introduce the idiom
    ``\oCODE{let (KMC  $\ldots$) = $\ldots$}''
    which enforces the type of the hole to be {\em monomorphic}.
    Otherwise, the type would be too general and
    this would spoil the type propagation (See \S~\ref{sec:instrumented}).
    % JL: sorry if i am dumbing this down too much!
    % 
    % generalises types like {\em any} to be {\em polymorphic}
    % $\forall \alpha.\alpha$ (can be instantiated with any type at {\em
    %   any site}) if its occurrence is {\em covariant}, spoiling the
    % propagation.  We avoid this by wrapping the pattern with a type
    % \oCODE{KMC : 'a -> 'a tuple} declared explicitly as non-covariant.
    % 
  \item The inferred type is translated to a system of
    (local) {\em session types}, which are
    passed to the \kmc-checker.%  that verifies whether the system
    % is \kmc.
    % 
  \item If the system is \kmc, then it is safe and the plugin {\em
      instruments} the code to allocate a fresh channel tuple 
      (i.e., concurrent queues) at the hole.
    % 
    % Figure~\ref{fig:instrumented} shows the instrumented code for
    % Figure~\ref{fig:fibworker}.
    %
    % In Figure~\ref{fig:instrumented}, Line~\ref{line:allocchan}
    % allocates {\em raw} channels using \oCODE{Chan.create_unbounded}
    % from Multicore OCaml, and
    % Lines~\ref{line:allocuch}-\ref{line:allocwch} create objects
    % inhabiting the inferred type.  We use shorthand
    %   \oCODE{<...>} for {\em in-place} objects \protect
    %   \oCODE{object...end} and abbreviate the continuations as
    %   ``\oCODE{...}''.
    %   % 
    %   The \oCODE{Internal} functions make a channel from raw channels
    %   and a continuation.  In particular, \oCODE{make_out} takes an
    %   extra function \oCODE{(fun v -> `label v)}, allocating a variant
    %   tag representing the message label.
    \item If the system is unsafe, the \kmc-checker returns a {\em
        violation trace} which is translated back to an OCaml type and
      inserted at the hole, to report a more precise error location.
\end{enumerate}

The translation is limited inside the \oCODE{[%kmc.gen]}
  expression, retaining a clear correspondence between the original
  and translated code.
  It can be understood as a form of {\em ad hoc polymorphism}
  reminiscent of type classes in Haskell.
  Like the Haskell typechecker verifies whether a type belongs to a
  class or not, the {\tt kmclib} verifies whether the set of session
  types belongs to the class of $k$-MC systems.

  \begin{figure}[t!]
    \begin{center}
            \includegraphics[width=9.5cm]{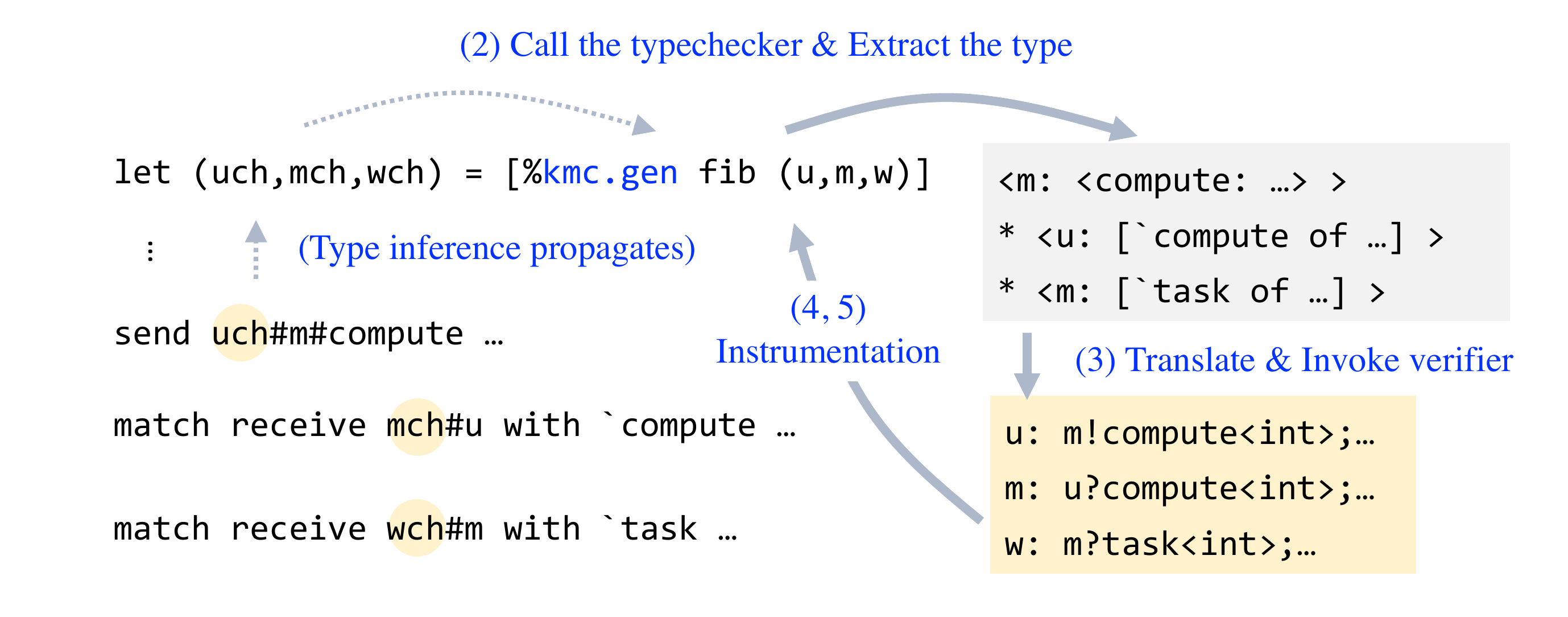}
          \caption{Inferring session types from OCaml code.\label{fig:inference}}
  \end{center}
  \vspace{-5mm}
\end{figure}

%%% Local Variables:
%%% mode: latex
%%% TeX-master: "main"
%%% End:

%  \input{kmc}
\section{Conclusion}
We have developed a practical library for 
safe message-passing programming. 
The library enables developers to program and 
verify arbitrary communication patterns
without the need for type annotations or user-operated external tools.
Our \emph{automated verification} approach can be applied to other
general-purpose programming languages.
Indeed it mainly relies on two ingredients: structural typing and
metaprogramming facilities. Both are available, with a varying degree
of support, in, e.g., Scala, Haskell, TypeScript, and F\#.

Our work is reminiscent of automated software model checking which has a
long history (see~\cite{10.1145/1592434.1592438} for a survey). There
are few works on inference and verification of behavioural types,
i.e.,~\cite{PereraLG16,LangeNTY17,LNTY2018,ASE21}.
However, Perera et al.~\cite{PereraLG16} only present a prototype
research language, while Lange et al.~\cite{LangeNTY17,LNTY2018,ASE21}
propose verification procedures for Go programs that rely on external
tools which are not integrated with the language nor its type system.
To our knowledge, ours is the first implementation of type inference
for MPST and the first integration of session types compatibility
checking within a programming language.

% This library is a basis for several future developments.   
% As a start, 

% perfect vehicle
% Due to space reasons, we have omitted several library
% features such as scatter/gather communication patterns,
% bound-aware transport optimisation, and static linearity checking. 

% Cover (MP)ST works and categorise them? 
% \begin{itemize}
% \item Scribble based
% \item Global combinators
% \item Binary ST works?
% \end{itemize}

% Future work:
% \begin{itemize}
% \item Prospect for other lgge
% \item other model checker (just have to agree input / counter-examples)
% \item transport
% \item Scatter / gather communication (extension of kmc)
% \end{itemize}

% We presented the most awesome tool you have ever seen. You are welcome. 
% %%% Local Variables:
% %%% mode: latex
% %%% TeX-master: "main"
% %%% End:

\bibliographystyle{abbrv}
\bibliography{jrk}

\begin{thebibliography}{10}

\bibitem{cfsm83}
D.~Brand and P.~Zafiropulo.
\newblock On communicating finite-state machines.
\newblock {\em J. {ACM}}, 30(2):323--342, 1983.

\bibitem{CastroHJNY19}
D.~Castro, R.~Hu, S.~Jongmans, N.~Ng, and N.~Yoshida.
\newblock Distributed programming using role-parametric session types in {Go}:
  statically-typed endpoint apis for dynamically-instantiated communication
  structures.
\newblock {\em {PACMPL}}, 3({POPL}):29:1--29:30, 2019.

\bibitem{ASE21}
N.~Dilley and {Julien Lange}.
\newblock Automated verification of {Go} programs via bounded model checking.
\newblock In {\em International Conference on Automated Software Engineering
  ({ASE})}. IEEE/ACM, July 2021.
\newblock To appear.

\bibitem{DBLP:conf/flops/Garrigue04}
J.~Garrigue.
\newblock Relaxing the value restriction.
\newblock In Y.~Kameyama and P.~J. Stuckey, editors, {\em Functional and Logic
  Programming, 7th International Symposium, {FLOPS} 2004, Nara, Japan, April
  7-9, 2004, Proceedings}, volume 2998 of {\em Lecture Notes in Computer
  Science}, pages 196--213. Springer, 2004.

\bibitem{harveyMultiparty2021}
P.~Harvey, S.~Fowler, O.~Dardha, and S.~J.~Gay.
\newblock Multiparty {Session} {Types} for {Safe} {Runtime} {Adaptation} in an
  {Actor} {Language}.
\newblock In A.~M{\o}ller and M.~Sridharan, editors, {\em 35th European
  Conference on Object-Oriented Programming (ECOOP 2021)}, volume 194 of {\em
  Leibniz International Proceedings in Informatics (LIPIcs)}, page~30,
  Dagstuhl, Germany, 2021. Schloss Dagstuhl -- Leibniz-Zentrum f{\"u}r
  Informatik.

\bibitem{HondaYC08}
K.~Honda, N.~Yoshida, and M.~Carbone.
\newblock Multiparty asynchronous session types.
\newblock In {\em {POPL} 2008}, pages 273--284, 2008.

\bibitem{HuY16}
R.~Hu and N.~Yoshida.
\newblock Hybrid session verification through endpoint {API} generation.
\newblock In {\em {FASE} 2016}, pages 401--418, 2016.

\bibitem{ImaiNYY19}
K.~Imai, R.~Neykova, N.~Yoshida, and S.~Yuen.
\newblock Multiparty session programming with global protocol combinators.
\newblock In {\em {ECOOP}}, volume 166 of {\em LIPIcs}, pages 9:1--9:30.
  Schloss Dagstuhl - Leibniz-Zentrum f{\"{u}}r Informatik, 2020.

\bibitem{10.1145/1592434.1592438}
R.~Jhala and R.~Majumdar.
\newblock Software model checking.
\newblock {\em ACM Comput. Surv.}, 41(4), Oct. 2009.

\bibitem{KouzapasDPG16}
D.~Kouzapas, O.~Dardha, R.~Perera, and S.~J. Gay.
\newblock Typechecking protocols with {Mungo} and {StMungo}.
\newblock In {\em PPDP 2016}, pages 146--159, 2016.

\bibitem{LangeNTY17}
J.~Lange, N.~Ng, B.~Toninho, and N.~Yoshida.
\newblock Fencing off {Go}: liveness and safety for channel-based programming.
\newblock In {\em {POPL} 2017}, pages 748--761, 2017.

\bibitem{LNTY2018}
J.~Lange, N.~Ng, B.~Toninho, and N.~Yoshida.
\newblock A static verification framework for message passing in {Go} using
  behavioural types.
\newblock In {\em {ICSE} 2018}. ACM, 2018.

\bibitem{LangeY19}
J.~Lange and N.~Yoshida.
\newblock Verifying asynchronous interactions via communicating session
  automata.
\newblock In {\em {CAV} {(1)}}, volume 11561 of {\em Lecture Notes in Computer
  Science}, pages 97--117. Springer, 2019.

\bibitem{miuGenerating2020}
A.~Miu, F.~Ferreira, N.~Yoshida, and F.~Zhou.
\newblock Generating {Interactive} {WebSocket} {Applications} in {TypeScript}.
\newblock {\em Electronic Proceedings in Theoretical Computer Science},
  314:12--22, Apr. 2020.

\bibitem{NHYA2018}
R.~Neykova, R.~Hu, N.~Yoshida, and F.~Abdeljallal.
\newblock {A Session Type Provider: Compile-time API Generation for Distributed
  Protocols with Interaction Refinements in F$\sharp$}.
\newblock In {\em {CC} 2018}. ACM, 2018.

\bibitem{NgCY15}
N.~Ng, J.~G. de~Figueiredo~Coutinho, and N.~Yoshida.
\newblock Protocols by default - safe {MPI} code generation based on session
  types.
\newblock In {\em {CC} 2015}, pages 212--232, 2015.

\bibitem{DBLP:journals/jfp/Padovani17}
L.~Padovani.
\newblock A simple library implementation of binary sessions.
\newblock {\em J. Funct. Program.}, 27:e4, 2017.

\bibitem{PereraLG16}
R.~Perera, J.~Lange, and S.~J. Gay.
\newblock Multiparty compatibility for concurrent objects.
\newblock In {\em {PLACES} 2016}, pages 73--82, 2016.

\bibitem{ScalasDHY17}
A.~Scalas, O.~Dardha, R.~Hu, and N.~Yoshida.
\newblock A linear decomposition of multiparty sessions for safe distributed
  programming.
\newblock In {\em {ECOOP} 2017}, pages 24:1--24:31, 2017.

\bibitem{DBLP:journals/pacmpl/Sivaramakrishnan20}
K.~C. Sivaramakrishnan, S.~Dolan, L.~White, S.~Jaffer, T.~Kelly, A.~Sahoo,
  S.~Parimala, A.~Dhiman, and A.~Madhavapeddy.
\newblock Retrofitting parallelism onto ocaml.
\newblock {\em Proc. {ACM} Program. Lang.}, 4({ICFP}):113:1--113:30, 2020.

\bibitem{YHNN2013}
N.~Yoshida, R.~Hu, R.~Neykova, and N.~Ng.
\newblock The {Scribble} protocol language.
\newblock In {\em TGC 2013}, volume 8358, pages 22--41. Springer, 2013.

\bibitem{DBLP:journals/pacmpl/00020HNY20}
F.~Zhou, F.~Ferreira, R.~Hu, R.~Neykova, and N.~Yoshida.
\newblock Statically verified refinements for multiparty protocols.
\newblock {\em Proc. {ACM} Program. Lang.}, 4({OOPSLA}):148:1--148:30, 2020.

\end{thebibliography}

\newpage
\appendix
\appendix

\section{Technical Details on the \kmclib API}
\label{sec:appimpl}

% We give a detailed account of the design and implementation of our library.
% We give an overview of some technical details of our library. 
%Two insihghts underpin the development of \kmclib -- 
%meta-programming facilities and structural typing.  
%
%Figure~\ref{fig:workflow} gives an overview of its architecture.
%

% \begin{itemize}
% \item structurally typed object
% \item Row polymorphism
% \item  PPX
% \end{itemize}

% \subsection{The \kmclib{} API}\label{subsec:kmclib}
%
% \textbf{The \kmclib{} API} 
% The \kmclib{} primitives allow the vanilla
% OCaml typechecker to infer the session structure of a program, while
% simultaneously providing a consistent communication API for the
% programmer. To enable inference of session types from communication programs,
% we leverage OCamls structural typing and row polymorphism.
% In particular, we reuse the encoding from \cite{ImaiNYY19}
% where input and outputs session types
% are encoded as polymorphic variants and objects in OCaml.
% Differently than \cite{ImaiNYY19} which relies on a top-down
% approach, \kmclib{} enables
% session channels to be inferred and verified without the need of a global type. 
We explain the main communication primitives of \kmclib{} and their translation
to session types. In particular, we reuse the encoding~\cite{ImaiNYY19} 
where input and output session types are encoded as polymorphic variants and objects, 
while loops are naturally handled using {\em equi-recursive types} in OCaml.

Objects and variants in OCaml are structurally typed, which enables the
creation of ad-hoc types.
This allows the channel structure to be derived
from the usage of channels in \oCODE{send} and \oCODE{receive} primitives.  
% , e.g objects types are
% inferred automatically from the method that is invoked on them.
% We leverage structural typing and row polimorphism to   

\myparagraph{Output types} Sending a message, e.g., in
Line~\ref{line:usersend} of Figure~\ref{fig:fibworker}, is parsed as
\oCODE{send (uch#m#compute) 42} where the two chained {\em method
  calls} yields a port for sending \oCODE{compute} label to role \oCODE{m},
which in turn is passed to \oCODE{send} (together with a payload).
This corresponds to an {\em internal choice} where the \oCODE{user}
specifies a destination for its message and chooses a label from those
offered by the receiver.

The inferred type of channel object \oCODE{uch} is a nested object
type of the form \oCODE{<m: <compute: (int, $\chan$)> out>} where \oCODE{m} is
a method that returns an object that itself provides method
\oCODE{compute} (which returns a port for sending an \oCODE{int} payload and yielding a
continuation channel \oCODE{$\chan$}).
Note that the implementation of these methods is not provided
explicitly by the API nor the programmer, instead they are constructed
on-demand when invoking \oCODE{uch#m#compute}; i.e., objects are
generated automatically according to the method types that is invoked on them
(\oCODE{#} denotes method invocation).
Such object types correspond to session types of the form
{\color{brown}\tt m!compute<int>;$\chan$} (the translation is trivial).

% having method names with return types reflects the structure as\footnote{
% For brevity, we omit ellipsis \oCODE{..} inferred by the typechecker, providing more flexibility.
% }
% \oCODE{<m: <compute: (int, $k$) out> >}
% where \oCODE{<$\ldots$>} is a nested {\em object type},
% with payload \oCODE{int} and {\em continuation} $k$,
% which is tranlated into session type {\color{brown}\tt m!compute<int>;$k$} (an output of message {\tt compute} with payload type {\tt int} to {\tt m}) in the later pass.
%% of the latter is determined by the following usage of \oCODE{uch} re-bound in the same line.

% 
\myparagraph{Input types} To receive messages, as in
Lines~\ref{line:usrrcv}-\ref{line:usrrcvend} of Figure~\ref{fig:fibworker}, we use
\oCODE{uch#m} to return a channel object which effectively corresponds
to a port originating from role \oCODE{m}.
This channel is then passed to the \oCODE{receive} primitive, which
returns a {\em polymorphic variant} on which one needs to pattern
match for expected messages.

The inferred type of \oCODE{wch}, which specifies the expected messages and their respective continuation, is 
\begin{center}
  \oCODE{<u: [`compute of int * $\chan'$ | `stop of unit * $\chan''$] inp>}
\end{center}
This type corresponds to an external choice, i.e., session
type of the form 
\begin{center}
  {\color{brown}\tt \{u?compute<int>;$\chan'$\} or
  \{u?stop<unit>;$\chan''$\}}.
\end{center}

% 
% \noindent\textbf{Recursion} Loops are naturally handled using {\em equi-recursive types} in OCaml.
%
% (\oCODE{$t$ as 'a}), which automatically unfolds to $t$ with variable
% \oCODE{'a} replaced with (\oCODE{$t$ as 'a}) itself.
%
% Finally, the \oCODE{close} requires the channel having \oCODE{unit}
% type, without any further communication on that port.

\myparagraph{Linearity}
MPST require channels to be used linearly,
i.e., each channel
must be used exactly once. 
If a channel is not used, this leads to a multiparty compatibility
issue (a message will not be sent/received), and hence our
implementation detects such issues statically via \kmc.

The idiomatic {\em shadowing} with the same variable names (e.g.,
re-binding of \oCODE{uch} in Line~\ref{line:usersend}) in OCaml
mitigates the risk of using a channel more than once.
If the program deviates from this best practice and a channel is
used non-linearly, an exception is raised at runtime.

Alternatively, \kmclib{} provides an
event-based alternative API (similar to that
of~\cite{DBLP:journals/pacmpl/00020HNY20}), which eliminates the
explicit need for linear channels, at the cost of losing a
direct-style API.\footnote{See \url{https://github.com/keigoi/kmclib/blob/tooldemo/test/paper/test_handler.ml} for an example.}
We remark that there are other known ways to check linearity
statically~\cite{ImaiNYY19}, which can easily be adapted to our library. 
%\input{workflow}

% \subsection{The Typed PPX Rewriter}\label{subsec:ppx}
%

\begin{figure}[t]
  \begin{center}
\begin{LISTING}[basicstyle=\scriptsize\CODESTYLE,numbers=left]
let KMC (uch,mch,wch) =
  let um, mu, mw, wm = Chan.create_unbounded (), Chan.create_unbounded (), ^\ldots^ in^\label{line:allocchan}^
  let uch = <m = <compute = Internal.make_out um (fun v -> `compute v) ^\ldots^ > > in^\label{line:allocuch}^
  let mch = <u = Internal.make_inp mu ^\ldots^ (fun v -> `compute v) > in^\label{line:allocmch}^
  let wch = <m = Internal.make_inp mw ^\ldots^ (fun v -> `stop v) > in^\label{line:allocwch}^
  make_tuple (uch, mch, wch)^\label{line:maketuple}^
\end{LISTING}
  \end{center}    
  \caption{Code from Figure~\ref{fig:fibworker}, instrumented at \oCODE{[kmc.gen]}}\label{fig:instrumented}
\end{figure}

%%% Local Variables:
%%% mode: latex
%%% TeX-master: "main"
%%% End:

\section{Instrumented Code for Figure~\ref{fig:fibworker}}
\label{sec:instrumented}

    Figure~\ref{fig:instrumented} shows the instrumented code for
    Figure~\ref{fig:fibworker}.
    Line~\ref{line:allocchan} allocates {\em raw} channels using
    \oCODE{Chan.create_unbounded} from Multicore OCaml, and
    Lines~\ref{line:allocuch}-\ref{line:allocwch} create objects
    inhabiting the inferred type.  We use shorthand \oCODE{<...>} for
    {\em in-place} objects \protect \oCODE{object...end} and
    abbreviate the continuations with an ellipsis.
      The \oCODE{Internal} functions make a channel from raw channels
      and a continuation.  In particular, \oCODE{make_out} takes an
      extra function \oCODE{(fun v -> `label v)}, allocating a variant
      tag representing the message label.  Also, it uses type casts
      from \oCODE{Obj} module in OCaml, which is a common technique to
      implement session types in OCaml
      (cf. \cite{DBLP:journals/jfp/Padovani17}).

      Line~\ref{line:maketuple} (\oCODE{make_tuple}) wraps the resulting
      tuple with the \oCODE{KMC} constructor.  As mentioned in
      \S~\ref{sec:impl}, this makes the inferred hole type {\em
        monomorphic}.  Normally, for the top-level declarations, OCaml
      generalises the type to be {\em polymorphic} and the hole type
      is inferred as $\forall \alpha.\alpha$ (can be instantiated with
      any type at {\em any site}) if its occurrence is at the {\em
        covariant} position, spoiling the propagation (cf. relaxed
      value restriction \cite{DBLP:conf/flops/Garrigue04}).  We avoid
      this by wrapping the pattern with a type \oCODE{KMC : 'a -> 'a tuple}
      declared explicitly as non-covariant.
    % JL: sorry if i am dumbing this down too much!
    % 
      
%   \item If the system is unsafe, a {\em violation trace} from
%       \kmc-checker is translated back to the OCaml type and inserted
%     at the hole, to report the precise location of the errors.
% \end{enumerate}

% The translation is limited inside the \oCODE{[%kmc.gen]}
% expression, retaining clear correspondence between the original and translated code.
% It is understood as a form of {\em ad hoc polymoprhism} reminiscent of  type classes in Haskell;
% whereas a Haskell type class checks a type belongs to a class or not;
% the {\tt kmclib} verifies whether the set of session types belongs to the class of $k$-MC system.

% \begin{figure}[t!]
%   \begin{center}
%     \includegraphics[width=10cm]{figures/inference.pdf}
%     \caption{Inferring session types from the code.\label{fig:inference}}
%   \end{center}
% \end{figure}

\section{Error Reporting with Type Ascription}
\label{sec:errorreport}
%% The errors in \kmclib programs are reported as a type error during compilation process, via typed PPX.
It is vital to show the {\em location} of the error to the programmer
when an error is found.
To achieve this, the PPX plugin of \kmclib instruments an extra {\em ascription} of an {\em incompatible type}
at the erroneous usage of a channel.
For example, see the error at Line~\ref{line:stuck-progress} in the left of Figure~\ref{fig:error-progress},
where the PPX plugin assigns the variable \oCODE{mch}
a type \oCODE{[`progress_violation]} (a single variant constructor type whose name is \oCODE{progress_violation}),
as the \kmc-checker detects the input blocking forever at that point.
Since it is used as 
\oCODE{<u: [`result of int * $\cdots$] inp >}
denoting an input of \oCODE{result} with an \oCODE{int} from the \texttt{user},
the OCaml typechecker reports a type error.
    
For usability purposes, \kmclib detects another kind of error, which
we refer to as {\em format error}.
These errors happen when
% a {\em misuse} of the channel where
the inferred type of a channel is not even in the form of output or
input session type (channel {\em misuse}).
For example, if the one drops the role name (\oCODE{#m}) writing \oCODE{send uch#compute 42},
the variable \oCODE{uch} has the inferred type \oCODE{<compute: (int,$\cdots$) out>},
which is not a session type anymore.
Figure~\ref{fig:format-error} shows such an error.
The highlighted part is assigned a type \oCODE{[`shoud_be_inp_or_out_object]} type
saying that the expression needs another method call (or the expression should be used as an input).
We are planning to improve error messages to be more descriptive,
e.g., as \oCODE{[`role_or_label_not_given]}.

Note that these format checks are all done within the \oCODE{[%kmc.gen]}
  (or \oCODE{[%kmc.check]}) primitive.
    These errors could also be regarded as a ``no instance'' error in
    the type class, as such ill-formatted types are not in the class
    of $k$-MC systems (they are not even in the class of session type
    syntax).

\begin{figure}[t!]
  \begin{center}
    \includegraphics[width=8cm]{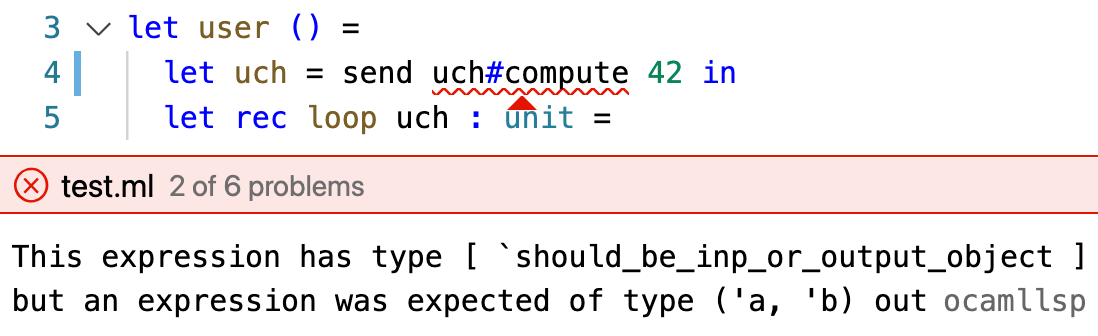}
  \end{center}
  \caption{Example of a format error.}\label{fig:format-error}
\end{figure}

%%% Local Variables:
%%% mode: latex
%%% TeX-master: "main"
%%% End:

\end{document}